# Spintronics-compatible approach to solving maximum satisfiability problems with probabilistic computing, invertible logic and parallel tempering

Andrea Grimaldi[1], Luis Sánchez-Tejerina[1,2], Navid Anjum Aadit[3], Stefano Chiappini[4], Mario Carpentieri[5], Kerem Camsari[3,*], Giovanni Finocchio[1,*]

[1]Department of Mathematical and Computer Sciences, Physical Sciences and Earth Sciences, University of Messina, 98166, Messina, Italy

[2]Department of Applied Physics, University of Salamanca, 37008, Salamanca, Spain

[3]Department of Electrical & Computer Engineering, University of California, Santa Barbara, CA 93106, Santa Barbara, United States

[4]Istituto Nazionale di Geofisica e Vulcanologia, Via di Vigna Murata 605, 00143 Roma, Italy

[5]Department of Electrical and Information Engineering, Politecnico di Bari, 70126 Bari, Italy

*Corresponding author: camsari@ucsb.edu, gfinocchio@unime.it

**Abstract**

The search of hardware-compatible strategies for solving NP-hard combinatorial optimization problems (COPs) is an important challenge of today's computing research because of their wide range of applications in real world optimization problems. Here, we introduce an unconventional scalable approach to face maximum satisfiability problems (Max-SAT) which combines probabilistic computing with p-bits, parallel tempering, and the concept of invertible logic gates. We theoretically show the spintronic implementation of this approach based on a coupled set of Landau-Lifshitz-Gilbert equations, showing a potential path for energy efficient and very fast (p-bits exhibiting ns time scale switching) architecture for the solution of COPs. The algorithm is benchmarked with hard Max-SAT instances from the 2016 Max-SAT competition (e.g., HG-4SAT-V150-C1350-1.cnf which can be described with 2851 p-bits), including weighted Max-SAT and Max-Cut problems.

**Introduction**

Combinatorial optimization problems (COPs) are a class of mathematical problems that have important applications in a variety of industrial and scientific fields, which span from logistics [1] to geoscience [2], from water distribution network design [3,4] to job scheduling [5]. Many of these, such as maximum cut (Max-Cut), Boolean satisfiability (SAT) or the travelling salesman problem, are NP-complete or NP-hard, meaning in their worst-case instances they have no polynomial-time solution. This makes them very hard, in terms of computational time, to solve when scaled to large sizes with conventional algorithms [6,7] designed to work on a von Neumann architecture. For this reason, unconventional approaches based on physical models, which can be used as paradigms that aim at rethinking the overall structure of the architecture in addition to devising different algorithms, have been investigated in the last years [8–13].

In particular, Ising machines (IMs), a computing paradigm based on the Ising model, are a promising candidate for facing COPs due to their computationally friendly discretized nature and the robustness of their energy-minimization process.

An Ising model consists of a *d*-dimensional lattice in which each site is characterized by a discrete variable $m_i$, with $m_i \in \{-1, +1\}$, named "spin". Each $i^{th}$ site is biased by an external field $h_i$ and interacts with the $j^{th}$ site via an exchange interaction coefficient $J_{ij}$. The Hamiltonian of such a system is:

$$H(\mathbf{m}) = -\sum_{i,j} J_{ij} m_i m_j - \sum_i h_i m_i \qquad (1)$$

Originally developed as a toy model to describe ferromagnetism [14], the Ising model has been shown to represent general COPs that are encoded in the *J*-matrix and the *h*-vector. Ising Machines [15] (IM) are dedicated computers that can solve such COPs encoded as Ising models. From an implementation point of view, the IMs can be realized with quantum annealing [16], oscillators [17], coherent IMs [18,19], simulated bifurcation [20] and with probabilistic circuits [21–23]. Here, we will focus on the latter approach, where the discrete variable $m_i$ is the state of a p-bit [23], namely, a bistable stochastic element. The use of p-bits to implement IMs is also known as probabilistic computing (PC) which is a probabilistic graph model and can be also used to also represent directed networks such as Bayesian Networks [24]. PC has been already used for different applications [23], including the integer factorization problem [23], [25]. A p-bit state is given by:

$$m_i = sgn\{rand(-1, +1) + tanh(I_i)\} \qquad (2)$$

where *rand* and *sgn* are a function generating uniform random numbers between $-1$ and $+1$ and the sign function, respectively, and *tanh* is the hyperbolic tangent function. $I_i$ is the p-bit control signal, determined by the local bias $h_i$ and the interaction with the other p-bits via the *J* matrix:

$$I_i = I_0\big(h_i + \sum_j J_{ij} m_j\big). \tag{3}$$

$I_0$ is the coupling parameter that acts as an inverse temperature, meaning that $I_0 = 0$ corresponds to infinite temperature and it is the parameter used to set and control the annealing process. Note that in the text, we will use both terms IM and PC to refer to the solution of Eqs. 2 and 3 based on interacting p-bits.

This work expands the use of PC to the solution of maximum satisfiability problems (Max-SAT), which have many industrial [26] and fundamental [27] applications, and how it can be generalized for solving weighted Max-SAT and Max-Cut problems. In fact, because of the importance and complexity of this category of problems, a SAT competition has been established in 2002 and is held every year to test state-of-the-art solvers on various instances of satisfiability problems [26]. Simulations point out that it is possible to face those problems with simulated annealing. Here, we show an annealing algorithm based on parallel tempering (PT). This approach adds redundance enhancing the capability of the IM state to escape from local minima.

We wish to stress that the main aim of this work is not to compare the PC-based approach with other implementations [28] of Max-SAT solvers such as D-Wave Advantage quantum annealers [29], continuous-time analog solvers [27,30], self-organizing memcomputing logic gates [31] and simulated bifurcation machines [20]. On the other hand, we provide a path to introduce spintronic technology as a platform for facing COPs.

In fact, the interest on PC is also exploding because of its compact hardware implementation with spintronic technology [21], taking advantage of the probabilistic nature of the switching between two states in superparamagnetic magnetic tunnel junctions (MTJs). Currently, this research direction is very active with the recent demonstration of nanosecond generation of p-bits by Tohoku University [32] and IBM [33]. In addition, spintronic-based PC has been used with success to experimentally realize invertible logic gates and to solve integer factorizations problems of numbers described by 8 p-bits [25]. In this context, we show the performance of a software implementation of PC based on a coupled set of Landau-Lifshitz-Gilbert (LLG) equations, one for each p-bit, in solving the Max-SAT instance "s3v70c700-1.cnf" from 2016 Max-SAT competition. The instance has 70 variables and 700 clauses and is encoded with 771 p-bits. Numerical simulations predict that for this problem the optimal solution can be reached in less than 60 ns if p-bits are implemented directly in hardware with MTJs. This would potentially allow the development of a spintronic solver orders of magnitude faster than professional solvers that are used in digital computers, implemented in FPGA or ASICs. This work provides a direction for PC in solving COPs with potentially ultralow energy cost and high speed.

**Invertible logic and Max-SAT instance**

Max-SAT is a generalization of the Boolean satisfiability problem and, as such, consists of a Boolean formula in conjunctive normal form which can be converted in a Boolean circuit realized with AND and OR logic gates. The implementation using PC makes use of the invertible counterpart of these logic gates. In detail, the use of invertible logic gates gives a way to map some combinatorial problems which can be expressed in term of logic gates in the PC paradigm ($J$ and $h$ Ising elements). The fact that those invertible gates can work in 'inverse' mode is the fundamental characteristic driving the exploration of the solution in the search space. This means that, if an entire circuit is composed of such gates, it can be used in reverse by clamping the output bits of the circuit, allowing the inputs to float between all the configurations that are compatible with the clamped values. [23] The goal of Max-SAT is to satisfy as many clauses as possible or, more accurately, to decrease the "cost" (the sum of the weights of the unsatisfied clauses) of the solution as much as possible. This can be seen as an energy minimization process, whose absolute minimum value is defined as the 'optimal' solution cost.

Thus, before calculating the solution using this IM, (i) the regular gates in the circuit are replaced with their equivalent invertible gate, (ii) the Max-SAT instance is then mapped into exchange and field matrices of the Ising model of Eq. 1 according to the topology of the logic circuit, and (iii) the p-bits linked to the output variables are clamped. Those steps have been already proposed in realizing logic circuits with invertible logic gates for the solution of integer factorization with PC [23]. A similar approach was pioneered by Di Ventra and Traversa for self-organizing logic gates in the digital memcomputing [31,34]' [35] paradigm.

Figure 1(a) shows the logic circuit for a toy instance of Max-SAT, which is shown in its ".cnf" file format [36] in the top left. This format is the international standard way of formatting a Max-SAT instance. "cnf" stands for "conjunctive normal form", the logical scheme of a Max-SAT instance. Every line of a ".cnf" file that doesn't begin with a c or a p (comments or parameters) is a clause of the instance. Each number represents a variable and if it is preceded by a minus sign it means that in that clause the variable is negated. Each line has to be ended with a 0, which corresponds to no variable. In Figure 1(a) the instance has 4 clauses and 7 variables. On the top right the same instance is shown in literal proposition form. Each of the four clauses is a disjunction of literals where the variables are in an OR relation with each other, as one can see by the circuit scheme of the instance on the bottom of Figure 1(a). The matrix describing the invertible OR gate has a dimension of $(2n_{lc} - 1)$, $n_{lc}$ being the number of literals of the clause. This logic circuit is mapped in the exchange and field matrices as indicated in Figure 1(b), considering the AND and OR invertible gates developed in [23]. Each clause matrix constitutes a block of the complete exchange matrix $J$ indicated with a black square, the integer value assigned to the matrix element can be identified by the color scheme. All the blocks are in an AND relation and are thus connected with each other by an off-diagonal positive element (marked with an O in the figure). The connected p-bits are the outputs of the circuit and are clamped. Along with the exchange matrix, each clause is also characterized by a bias field vector $h$. The color bar in Fig. 1(b) is used to show the values of the matrix element, which are discrete, integer values amenable for efficient hardware implementation. This is a consequence of the composable nature of invertible logic circuits: the same AND, OR, NOT based building blocks are used to design composed circuits with the same integer weights of the building blocks.

Finally, because the same literal may appear in more than one clause, all the instances of a variable are connected with off-diagonal elements in the same way as the output p-bits (in the figure the letters from A to G are the connections for the variables 1 to 7). The number of connections is the number of permutations of the $n$ variables in a subset of 2, that is $\frac{n!}{(n-2)!} = (n-1)n$.

The PC software implementation based on Eqs. 2 and 3 works similarly to WalkSAT-based algorithms, because the p-bits are updated sequentially. However, as discussed ahead in the text, one potential strength of PC is its hardware implementation with spintronic technology, which will allow the parallel update of p-bits at hardware level being the p-bits described by a dynamical system. Mind that this characteristic cannot be directly transferred to a Field Programmable Gate Arrays (FPGA)-based implementation because the p-bit is realized with a digital random bit generator, although we recently proposed an approach to introduce some parallelism in FPGA-based PC implementation [37].

The Max-SAT instances solved in this work are from the Max-SAT 2016 competition. The standard problems are taken from the ms_random category, the Max-Cut instance is from the ms_crafted and the weighted instance from the wpms_random.

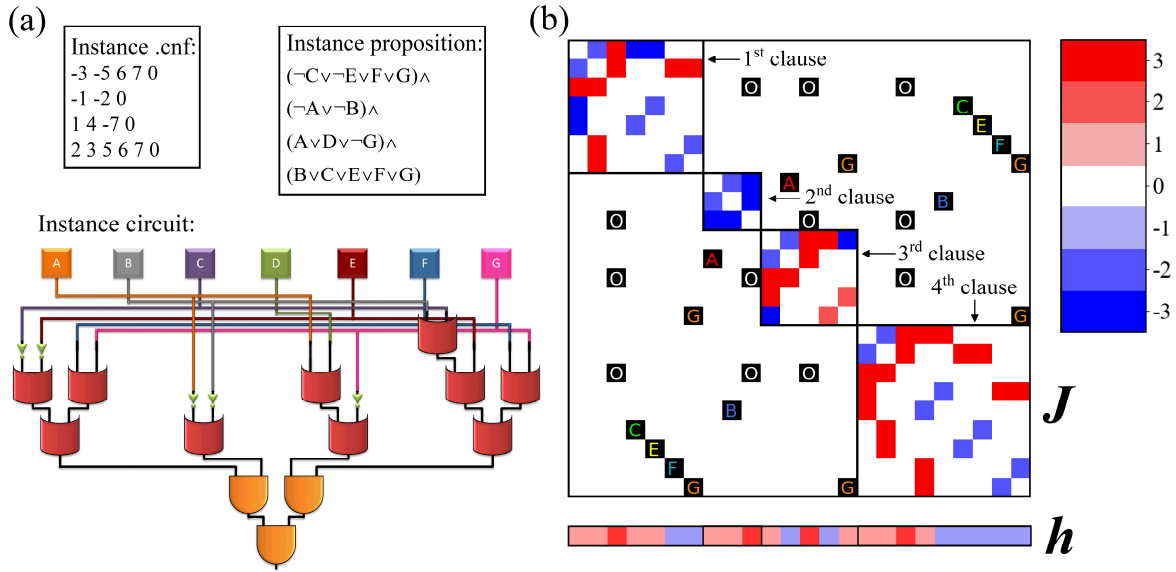

Figure 1 (a) A toy instance in its ".cnf" file format (top left), its logical proposition form where the letters from A to G are the variables from 1 to 7 (top right) and its circuit scheme (bottom). (b). The exchange matrix $J$ and bias field $h$ of the toy instance of Max-SAT shown in (a). The full interaction matrix takes the form of a block matrix where each block represents a clause of the instance. The connections between p-bits are described with off-diagonal positive non-zero elements in the interaction matrix (the black squares in the picture). The clauses are connected to each other by their output p-bits (their connections are marked with an O). The connection between p-bits that represent the same variable (letters A to G) is handled in the same way. The color bar is used to show the values of the matrix element.

**Annealing with parallel tempering**

To perform the calculations, we had to choose an annealing algorithm that drives the PC to minimize the number of unsatisfied clauses. One of the most popular approaches is the simulated annealing [38], however, its software implementation in PC is very difficult to parallelize because of the need of the sequential update of p-bits [23]. This aspect limits the possibility to speed up the calculations by using GPU-based algorithms. A possible directions to implement PC with parallel update of the p-bits is an annealing algorithm based on parallel tempering [39,40], a method devised for Monte Carlo simulations which uses a set of interacting replicas of the system at different temperatures.

Parallel tempering has already been used in its original formulation in IMs for Max-SAT [28], but has never been applied to PC. The computation of the state for each replica occurs in parallel at two different pseudo-temperatures, red (blue) line corresponds to high (low) temperature. Higher temperatures serve, in heuristic terms [41], as the diversification element of the algorithm: their highly stochastic behavior and strong fluctuations allow them to explore a larger ensemble of states in the search space, until they find a configuration with a lower energy than the neighbor replica at lower temperature. At this iteration a switching occurs, the state held by the high temperature replica is exchanged with that of the low temperature one. These replicas act as the intensification element: the thermal fluctuations are minimal and aim at improving the state as much as possible by exploring the local solution until a minimum is found. Because of the low stochasticity of this process, cold replicas are usually unable to get out of deep local metastable minima, so they require hotter replicas to find better starting states for them to minimize. Figure 2(a) shows an example illustrating the implementation of parallel tempering used here considering two replicas. Each starts with a state, in the example of Fig. 2(a) we named them as State 1 (continuous line) and State 2 (dashed line), evolving in the hotter and colder temperature replica respectively. If the replica energies, evaluated as the number of unsatisfied clauses in the case of Max-Sat problems, intersect at a given iteration, switching occurs (switching time in Fig. 2(a)) and State 1 is now evolving in the colder replica while State 2 in the hotter one.

In this annealing algorithm, the number of replicas and their temperatures are very important parameters to set. The ideal configuration has each replica exploring an ensemble of states that slightly overlaps with those of the hotter and colder replicas. Additionally, the coldest replica ought to have a low enough temperature so that it could not escape from any energy minimum. There is not an established procedure to identify the value of $I_0$ for each replica. Hence, we have used an empirical approach based on four main criteria:

1. Fast computation and low memory occupancy. We have chosen as few replicas as possible characterized by a vector of ordered $I_0$.

2. One replica is set at a low enough temperature to maintain a potential solution.

3. A replica explores the solution space of a problem with energy cost fluctuating around an average energy value and with a standard deviation that depends on the temperature value. The explored solution space for a given ith replica should overlap with that of its neighbors $(i+1)^{th}$ and $(i-1)^{th}$ replica.

4. The values of $I_0$ depend on the average number of p-bit neighbors and it is higher in networks with more interacting p-bits. This is necessary to have strong enough fluctuations to overcome the energy barrier set by the p-bits interaction of potential metastable states which is set by the p-bits interaction.

As evidenced, the choice of temperature depends on the topology of the problem. If the choice is not adequate, the increased computational cost of the additional replicas doesn't bring any improvement in the solving process. As an example, Fig. 3 shows a parallel tempering configuration set with a not effective $I_0$ vector according to the point 3 of the above list, for this computing scheme only the replica at $I_0=0.6$ works to find the optimal solution.

In summary, the optimal choice of the $I_0$ for each replica changes from problem to problem, depending on factors like sparsity of the connections, their strength, the average numbers of neighbors per p-bit, etc.

We have performed a systematic study finding that the use of four replicas, which has been fixed for the studies presented in the rest of this work, is sufficient to observe a speed-up in the convergency toward an energy minimum in the Hamiltonian of Eq. 1. The number is, for the problem considered, an adequate tradeoff between convergence speed and computational speed. More replicas would only slightly reduce the number of iterations required to get to the solution but considerably increase the computational cost of an iteration. The opposite would happen with less replicas. In performing a simple comparison between simulated annealing and parallel tempering, we have found that parallel tempering is comparable in the convergence velocity to an optimal solution in various small/medium size instances of Max-SAT problems (not shown).

Fig. 2(b) shows a performance test of parallel tempering for the Max-SAT instance "s3v70c700-1.cnf". PT has been tested by performing 100 solving trials and keeping the best solution found within $10^4$ iterations. The optimal (absolute minimum) number, for the instance considered, is 21. As one can see, the results show that parallel tempering manages to get to the optimal over 65% of the times, it reaches the optimal solution 100% of the times when considering $10^5$ iterations. The same test is performed with 5 replicas ($10^4$ iterations) and a comparison with the 4 replicas is shown in Fig. 4 (a). While we do see a noticeable improvement in the number of times the optimal is reached for that instance, from Fig. 4 (b) we can see that the difference in performance does not hold as the size and the difficulty of the instance increases.

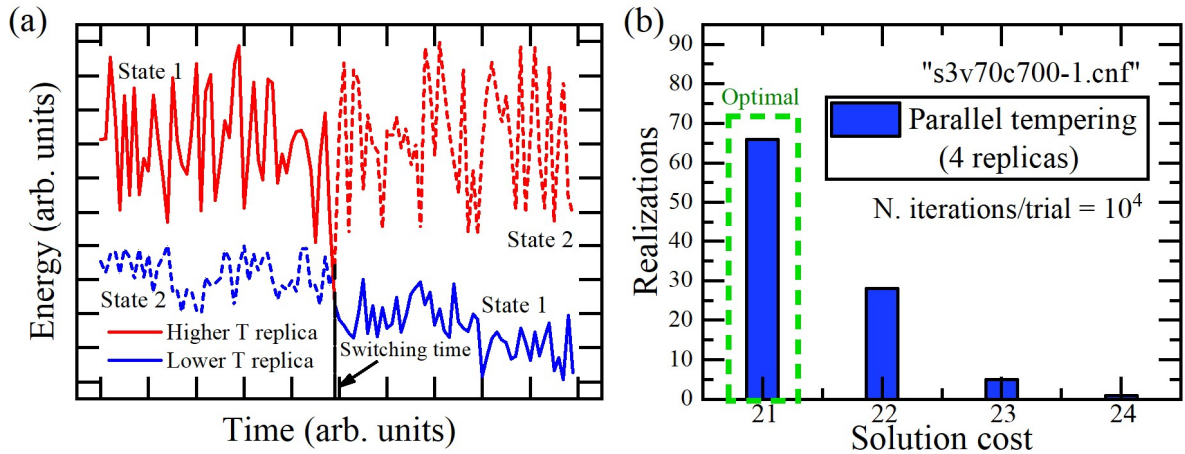

Figure 2. (a) A schematic representation of parallel tempering with two replicas at low and high temperature, blue and red lines, respectively. Replicas hold a state of the system (the continuous and dashed lines) and evolve it with the evolution algorithm. The State 1 switches with the State 2 at the iteration (switching time) when its energy become smaller than the State 2 energy. (b) Histogram showing the statistic of the final solution cost computed for 100 realizations for the Max-SAT instance "s3v70c700-1.cnf" (70 variables and 700 clauses, encoded with 771 p-bits).

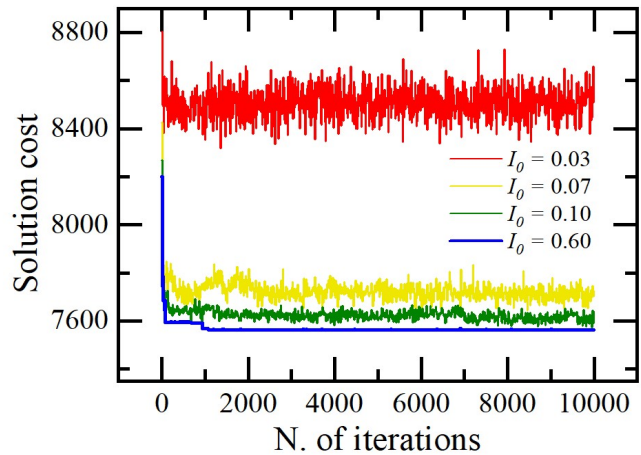

Figure 3. Solution of the instance "g1.rud" (800 variables and 38352 clauses, encoded with 800 p-bits) by using four replicas with the following $I_0$ vector: [0.03, 0.07, 0.10, 0.60]. Each colored line represents the evolution of the solution cost of a replica (0.03 in red, 0.07 in yellow, 0.10 in green, 0.60 in blue). As one can see, with this choice of temperatures, the replicas do not interact and, therefore, the only replica that has an impact on the calculation is the coldest one, with 75% of computational time wasted.

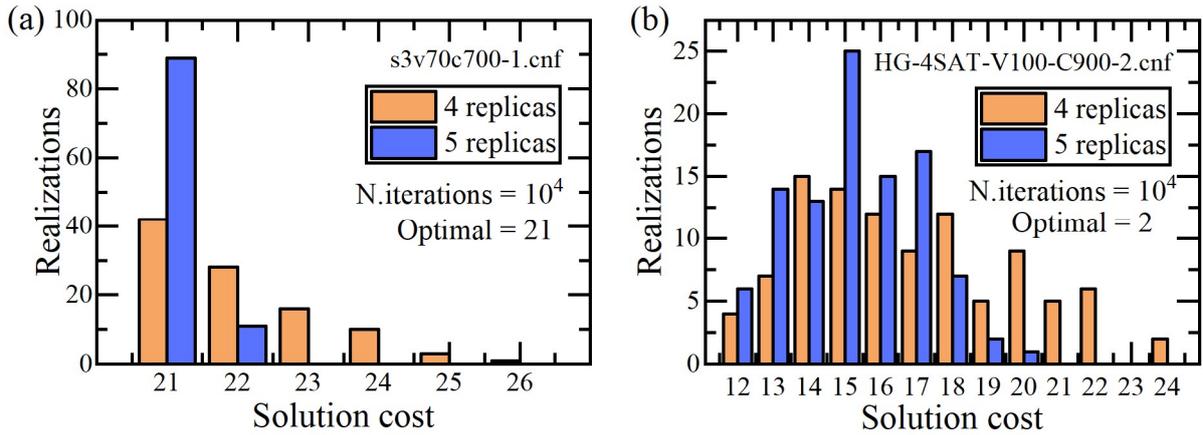

Figure 4. Histograms showing the statistics of the final solution cost computed for 100 realizations for the Max-SAT instances "s3v70c700-1.cnf" (70 variables and 700 clauses, encoded with 771 p-bits) in (a) and "HG-4SAT-V100-C900-2.cnf" (100 variables and 900 clauses, encoded with 1901 p-bits) in (b). In orange the results with parallel tempering with 4 replicas, in blue the ones with 5 replicas. All trials were made with 104 iterations. As one can see in (a), for the easier instance, 5 replicas perform strikingly better than 4, as the optimal is reached more than 80% of the time, compared to the 40% of the configuration with 4 replicas. However, in (b), the same two configurations performance are much closer. While with 5 replicas we can see slightly better results on average, the improvement is not enough to justify the 25% increase in computational time.

**Results and Discussion**

Here, we show the computational achievements for two hard Max-4SAT instances, "HG-4SAT-V100-C900-2.cnf" and "HG-4SAT-V150-C1350-1.cnf". 4SAT means that each clause contains 4 literals. The former instance is characterized by 900 clauses and 100 variables, a medium-hard instance in its category, which is solved in less than $6 \cdot 10^5$ iterations. The number of p-bits for such a system is 1901. The latter has 1350 clauses and 150 variables, one of the hardest instances in the competition, described by 2851 p-bits is brought to a nearly optimal solution (4 unsatisfied clauses instead of the optimal 1) in less than $8 \cdot 10^5$ iterations. The results are summarized in Figure 5 which shows an example of the time domain evolution of the cost function for 1 million of iterations (each replica has a different color), computed as the number of unsatisfied clauses for both instances. The pseudo-temperatures have been estimated with a systematic study of 1000 iterations trials so that the search space of a replica slightly overlaps that of the subsequent one, see for example the red and yellow curves. The pseudo-temperatures $I_0$ for each replica are indicated directly in the figures.

The last replica (the coldest one) is usually set at a value that makes the worsening of its state very unlikely. This is the control replica and is the one that usually reaches the optimal solution. An improvement of the solution cost is transferred at the colder replica as described in the previous paragraph.

We have included in the parallel tempering algorithm a "reset", in order to decrease the probability of getting stuck in local minimum. The reset inverts the states of each replica ($m_i = -m_i$), so that the new states are as distant as possible from the current ones in the search space. The reset occurs if the replica at the lowest temperature does not improve its state quality for more than a number of iterations. In Figure 5 (a) the threshold was set to $5 \cdot 10^4$ iterations, in Figure 5 (b) to $10^5$. The choice of the reset threshold is set empirically. As expected, we found that larger systems need a longer time before the reset schedule. The best solution is stored separately from the replicas, so that, even after a reset, the eventual optimum is not lost.

Longer simulations could potentially manage to get to the optimal configuration, but we have chosen to work here at a fixed number of iterations, equal to $10^6$, for the sake of comparison.

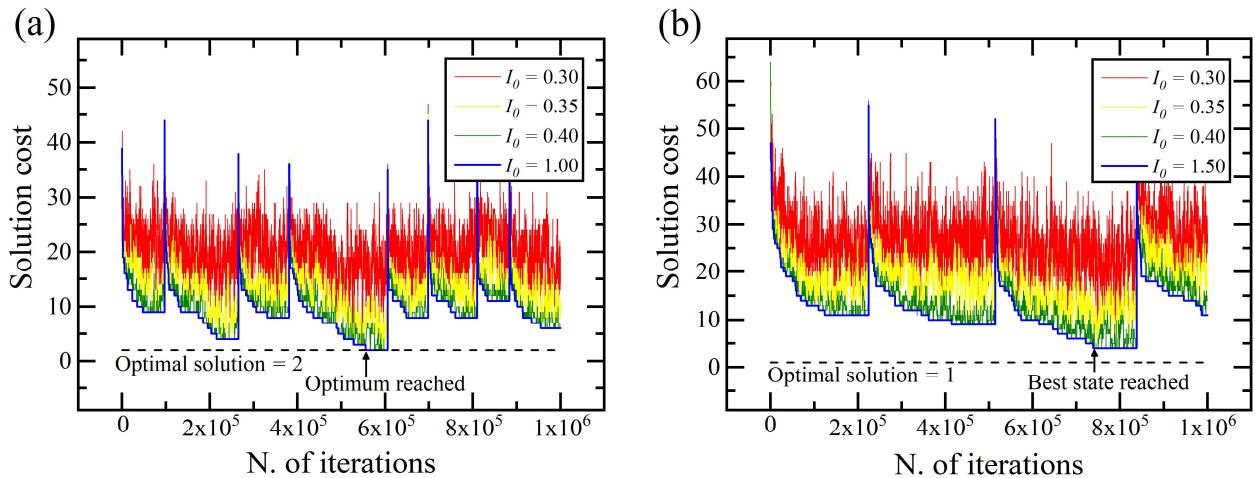

Figure 5. (a) Solution of the Max-SAT instance "MAXSAT_HG-4SAT-V100-C900-2.cnf" (100 variables and 900 clauses, encoded with 1901 p-bits) with the probabilistic solver. Four replicas are used, one with a substantially colder temperature in order to have a frozen-like state. When this replica remains stuck in the same state for a given number of iterations, all states are reset and the solver starts from scratch. In this case the optimal solution is 2 and the system obtains it at less than $6 \cdot 10^5$ iterations. (b) Solution of instance "HG-4SAT-V150-C1350-1.cnf" (150 variables and 1350 clauses, encoded with 2851 p-bits). The same methods as panel (a) is used, for this instance the optimum (1 in this instance) is not reached in 1 million of iterations, the results are still remarkable as the solver manages to get to a nearly optimal result. $I_0$ represents the pseudo-temperature for each replica.

Another benchmark has been performed for the Max-Cut problem which can be easily rewritten as a Max-2SAT instance by converting the edges between two nodes A and B into the logical proposition: $(A \lor B) \land (\neg A \lor \neg B)$. If the proposition is true, the edge is cut. Our solver can easily handle such problems, as shown in Figure 6 (a), where the optimal state of the spin-glasses Max-Cut instance "t7pm3-9999.spn.cnf" is found. The instance has 343 variables and 2058 clauses. Max-Cut is compatible with PC encoding, as it only requires 344 p-bits. Four replicas are used, one with a substantially colder temperature in order to have a frozen-like state. This instance has optimal $O$ equal to 209 and is reached by our solver in less the $10^4$ iterations. It should be noted the optimal does not correspond to the maximum cut, since each edge is represented by two clauses and both have to be satisfied in order to cut the connection. The maximum cut can be computed as $(N - 2O)/2$, where $N$ is the number of clauses and $O$ is the optimal value achieved by the PC. For this instance, then the resulting Max-Cut value is 820.

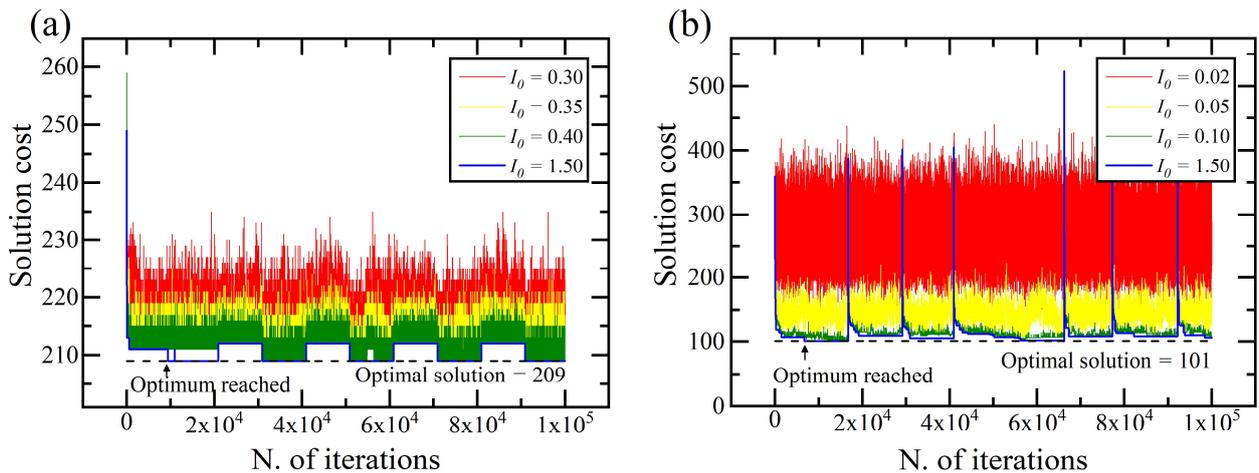

Figure 6. (a) Time domain evolution of the solution cost of the Max-Cut instance "t7pm3-9999.spn.cnf" (343 variables and 2058 clauses). The system is encoded with 344 p-bits. The optimal solution is reached in very few iterations (less than $10^4$). (b) Solving of the weighted Max-3SAT instance "s3v70c700-1.wcnf" (70 variables and 700 clauses) with the probabilistic solver. The number of p-bits here is 771. The weight is implemented naturally by multiplying each clause matrix and vector by its value.

Finally, we wish to highlight that Max-SAT can be further generalized by assigning a weight to each clause, meaning that some clauses are more important than others and their satisfaction should be prioritized. This weighted Max-SAT can be trivially implemented in our solver by properly scaling a proposition block matrix by its weight, thus intrinsically increasing its relevance. An example of this is shown in Figure 6(b), where a weighted instance, "s3v70c700-1.wcnf", from the Max-SAT 2016 competition is solved and its optimal is found. The instance has 70 variables and 700 weighted clauses. The encoding requires a total of 771 p-bits. Differently from an unweighted instance, the energy to minimize does not trivially coincide with the number of unsatisfied clauses, since each proposition has a different weight. Thus, we consider the solution cost –namely, the sum of the weights of the unsatisfied clauses. This instance has optimal equal to 101 and is reached by our solver in less the $10^4$ iterations. We wish to stress that convergence speed could be optimized further by exploring the effects of adding or removing replicas or by changing their temperatures. As the choice of the number of replicas and their temperatures was made after a human-interpreted, albeit rigorous, systematic study, advanced methods of parameters optimization could get to a different configuration that optimized the solving process further. Additionally, more advanced algorithmic strategies could also increase the accuracy and convergence speed of the solver. All these possibilities are being currently investigated for future works.

Here we wish to move the research on probabilistic computing forward by bringing in the game new annealing processes (such as the parallel tempering) and potential acceleration with a spintronic implementation of this paradigm (as shown in the following) with the aim to improve the TTS-100%.

**Spintronic based solution**

It has been already proved that it is possible to build these p-bits with spintronic technology [18] by using superparamagnetic MTJs [21]. This can be achieved by either reducing the size of an MTJ cross section [42] or by tuning the magnetic anisotropy properly by an external voltage. [43] Figure 7 (a) shows an example of time domain trace of a p-bit implemented with the numerical solution of the LLG equation within the macrospin approximation where the continuous jumps between two states of the MTJ occurs at nanosecond scale. [33, 44] Those two states code the binary information needed for the implementation of PC. The dynamical equation describing the single $i^{th}$ p-bit is given by [45,46]· [47]

$$\frac{d\mathbf{m}_i}{d\tau} = -\mathbf{m}_i \times \mathbf{h}_{\text{eff},i} + \alpha \mathbf{m}_i \times \frac{d\mathbf{m}_i}{d\tau} \qquad [4]$$

where $\mathbf{m}_i$ and $\mathbf{h}_{\text{eff},i}$ are the dimensionless magnetization and effective field of the free layer magnetization, $\alpha$ is the Gilbert damping and $d\tau$ is the dimensionless time $d\tau = \gamma_0 M_S dt$. The effective field takes into account the demagnetizing field $\boldsymbol{h}_{demag}$, the uniaxial anisotropy $\boldsymbol{h}_{anis}$, the thermal fields and the external field $h_i$ reported in the Eq. 1. The thermal field $\boldsymbol{h}_{th}$ is given by [48]

$$\boldsymbol{h}_{th} = \frac{\boldsymbol{\eta}}{M_s}\sqrt{\frac{2\alpha k_B T}{\gamma_0 \mu_0 M_s V dt}}, \qquad (5)$$

with $k_B$, $T$ and $V$ being the Boltzmann constant, the temperature and the computational system volume, respectively, while $\boldsymbol{\eta}$ is a vector whose Cartesian components are random numbers following the Gaussian distribution with a zero mean and unit variance [48], [49]:

$$\begin{cases} \langle \eta_k(t) \rangle = 0 \\ \langle \eta_k(t)\eta_l(t) \rangle = \delta_{kl}\delta(\boldsymbol{r}-\boldsymbol{r}')\delta(t-t') \end{cases}. \qquad (6)$$

The effective thermal field modulus is $3.8 \cdot 10^{-3} \cdot \sqrt{T}$. We want to stress that the temperature T in this case refers to the physical temperature of the device and is not correlated to the value of the pseudo-temperature $I_0$ of Equation (3), that is a parameter that scales the input signal of each p-bit to set the effective field that acts on the MTJ. The dimensionless effective anisotropy field is included in the demagnetizing field and it has the following expression $\boldsymbol{h}_{anis} + \boldsymbol{h}_{demag} = (-D_x m_x, -D_y m_y, -D_z m_z)$ where $D_x = -0.05$, $D_y = 0$, $D_z = 1$ are the effective demagnetizing tensors. [45], [50] The $D_x$ is negative because it takes into account the in-plane anisotropy along the x-direction (easy axis of the ellipse used to model the MTJ cross section). The Gilbert damping parameter is $\alpha = 0.1$ while gyromagnetic ratio $\gamma_0 = 2.21 \cdot 10^5$ m/(As). The time integration scheme used to solve the Eq. 4 is the 4$^{th}$ order Runge-Kutta.

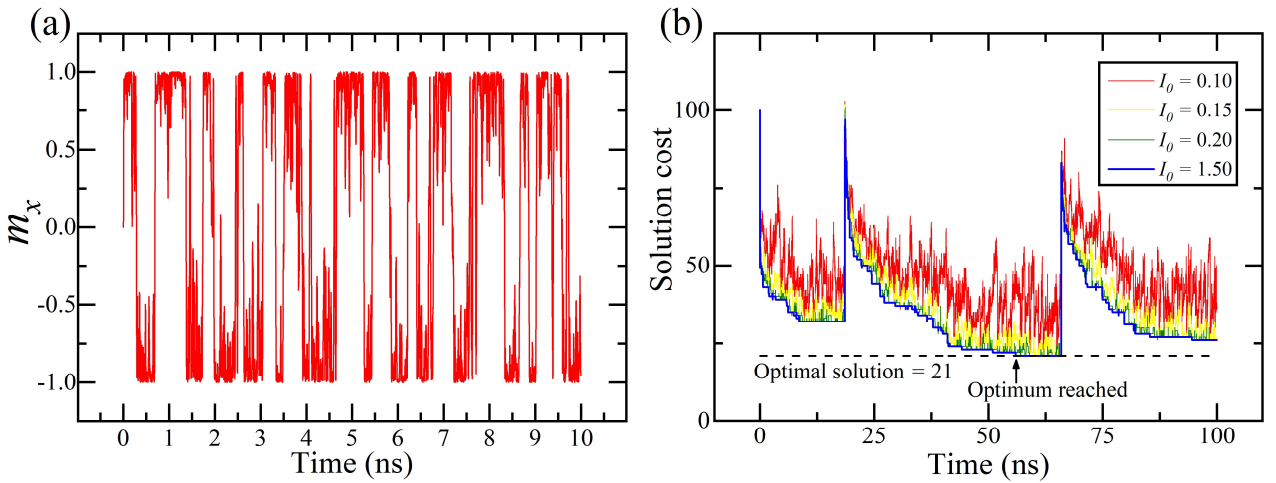

Figure 7. (a) Time domain traces of the three components of the magnetization dynamics of a superparamagnetic MTJ, the x-component is the one related to the p-bit dynamics. The simulation dynamics has been computed numerically with the LLG equation within the macrospin approximation. (b) Solution of the Max-3SAT instance "s3v70c700-1.cnf" (70 variables and 700 clauses) with the spintronic based PC solver. The number of p-bits for this instance is 771. The time evolution of the solution cost is obtained by integrating the LLG equation with a fourth order Runge-Kutta time integration solver. The time step used is $10^{-3}$ ns and the simulation was run for 100 ns. Four replicas are used, one with a substantially colder temperature in order to have a frozen-like state. When this replica remains stuck in the same state for a given number of iterations, all states are reset and the solver starts from scratch. In this case, the optimal solution is 21 and the system obtains it at less than 60 ns.

In this scenario, the Max-SAT instance is solved with the dynamical evolution of a number of macromagnetic equations equal to the number of p-bits, all those equations are coupled trough the exchange matrix derived for the problem to face. In other words, the coupling is modeled as an additional Zeeman field $I_i$ given by the Eq. 3 and added directly to the effective fields. Similar results are achievable by considering a three terminal device with the spin-orbit torque as proposed on [23] with a difference that the p-bit state is stabilized by the spin-orbit torque or in two terminal devices as shown in ref. [25]. In this dynamical approach the p-bit vector is updated in parallel, which is an advantage with respect to the model based on Eq. 2 where the p-bits need to be updated sequentially. [23] Figure 7 (b) shows the results achieved with this theoretical spintronic implementation of the PC considering the Max-SAT instance "s3v70c700-1.cnf". The simulation is long $10^5$ steps, which, converted in time units, correspond to 100 ns. The solver manages to get to the optimal state in less than 60 ns. This result which is a key finding of this work is a motivation to push the efforts in a future hardware implementation of PC with spintronic devices. We wish to highlight that this approach can be scalable being the electrical connections between different invertible logic gates the same as the ones used today for conventional CMOS circuits.

**Summary and Conclusion**

This work introduces a strategy for facing Max-SAT problems combining PC with parallel tempering. We solve small/medium instances of those problems reaching size 2851 p-bits (HG-4SAT-V150-

C1350-1.cnf) taken from 2016 Max-Sat competition showing that optimal or high-quality solution can be achieved with this approach.

The contributions of this work are two-fold: first, it extends the use of Invertible Logic in IMs to design sparse-graph representations for real-world MAX-SAT instances. Second, it reinforces the potential of CMOS-compatible spintronic technology that can achieve orders of magnitude speed-up in specialized, energy-efficient hardware. In particular, some of the authors presented a hardware implementation of PC with simple, non-optimized simulated annealing using FPGAs [37] which already have shown performance, in term of computational speed, capable of beating state-of-the-art solvers in the time required to reach 95% of the optimal solution (TTS-95%). We believe that the very recent promising results on PC will be of stimulus for the community to investigate further this research direction.

**Acknowledgements**

This work was supported under the project PRIN 2020LWPKH7 funded by the Italian Ministry of University and Research. G.F., A.G., L.S.T and M.C. would like to acknowledge the contribution of the COST Action CA17123 "Ultrafast opto-magneto-electronics for non-dissipative information technology". The work was also partially supported by PETASPIN association (www.petaspin.com).

**References**

[1]   M. R. Bartolacci, L. J. LeBlanc, Y. Kayikci, and T. A. Grossman, *Optimization Modeling for Logistics: Options and Implementations*, J. Bus. Logist. **33**, 118 (2012).

[2]   W. Zhang, C. Wu, H. Zhong, Y. Li, and L. Wang, *Prediction of Undrained Shear Strength Using Extreme Gradient Boosting and Random Forest Based on Bayesian Optimization*, Geosci. Front. **12**, 469 (2021).

[3]   H. M. V. Samani and A. Mottaghi, *Optimization of Water Distribution Networks Using Integer Linear Programming*, J. Hydraul. Eng. **132**, 501 (2006).

[4]   A. Krause, J. Leskovec, C. Guestrin, J. VanBriesen, and C. Faloutsos, *Efficient Sensor Placement Optimization for Securing Large Water Distribution Networks*, J. Water Resour. Plan. Manag. **134**, 516 (2008).

[5]  Z. Zhou, F. Li, H. Zhu, H. Xie, J. H. Abawajy, and M. U. Chowdhury, *An Improved Genetic Algorithm Using Greedy Strategy toward Task Scheduling Optimization in Cloud Environments*, Neural Comput. Appl. **32**, 1531 (2020).

[6]  M. R. Garey and D. S. Johnson, *Computers and Intractability : A Guide to the Theory of NP-Completeness* (1979).

[7]  H. R. Lewis, *Michael R. ΠGarey and David S. Johnson. Computers and Intractability. A Guide to the Theory of NP-Completeness. W. H. Freeman and Company, San Francisco1979, x + 338 Pp.*, J. Symb. Log. **48**, 498 (1983).

[8]  C. De Simone, M. Diehl, M. Jünger, P. Mutzel, G. Reinelt, and G. Rinaldi, *Exact Ground States of Ising Spin Glasses: New Experimental Results with a Branch-and-Cut Algorithm*, J. Stat. Phys. **80**, 487 (1995).

[9]  F. Liers, M. Jünger, G. Reinelt, and G. Rinaldi, *Computing Exact Ground States of Hard Ising Spin Glass Problems by Branch-and-Cut*, in *New Optimization Algorithms in Physics* (Wiley-VCH Verlag GmbH & Co. KGaA, Weinheim, FRG, 2005), pp. 47–69.

[10] M. Mézard, F. Ricci-Tersenghi, and R. Zecchina, *Two Solutions to Diluted P-Spin Models and XORSAT Problems*, J. Stat. Phys. **111**, 505 (2003).

[11] M. Mézard, G. Parisi, and R. Zecchina, *Analytic and Algorithmic Solution of Random Satisfiability Problems*, Science (80-. ). **297**, 812 (2002).

[12] M. Mézard and R. Zecchina, *Random K-Satisfiability Problem: From an Analytic Solution to an Efficient Algorithm*, Phys. Rev. E - Stat. Physics, Plasmas, Fluids, Relat. Interdiscip. Top. **66**, 27 (2002).

[13] A. P. Young, *Spin Glasses and Random Fields*, Vol. 12 (WORLD SCIENTIFIC, 1997).

[14] E. Ising, *Beitrag Zur Theorie Des Ferromagnetismus*, Zeitschrift Für Phys. **31**, 253 (1925).

[15] A. Lucas, *Ising Formulations of Many NP Problems*, Front. Phys. **2**, 1 (2014).

[16] A. B. Finnila, M. A. Gomez, C. Sebenik, C. Stenson, and J. D. Doll, *Quantum Annealing: A New Method for Minimizing Multidimensional Functions*, Chem. Phys. Lett. **219**, 343 (1994).

[17] J. Chou, S. Bramhavar, S. Ghosh, and W. Herzog, *Analog Coupled Oscillator Based Weighted Ising Machine*, Sci. Rep. **9**, 14786 (2019).

[18] T. Inagaki, Y. Haribara, K. Igarashi, T. Sonobe, S. Tamate, T. Honjo, A. Marandi, P. L.

McMahon, T. Umeki, K. Enbutsu, O. Tadanaga, H. Takenouchi, K. Aihara, K. Kawarabayashi, K. Inoue, S. Utsunomiya, and H. Takesue, *A Coherent Ising Machine for 2000-Node Optimization Problems*, Science (80-. ). **354**, 603 (2016).

[19] Y. Yamamoto, K. Aihara, T. Leleu, K. Kawarabayashi, S. Kako, M. Fejer, K. Inoue, and H. Takesue, *Coherent Ising Machines—Optical Neural Networks Operating at the Quantum Limit*, Npj Quantum Inf. **3**, 49 (2017).

[20] K. Tatsumura, M. Yamasaki, and H. Goto, *Scaling out Ising Machines Using a Multi-Chip Architecture for Simulated Bifurcation*, Nat. Electron. **4**, 208 (2021).

[21] K. Y. Camsari, S. Salahuddin, and S. Datta, *Implementing P-Bits With Embedded MTJ*, IEEE Electron Device Lett. **38**, 1767 (2017).

[22] R. Faria, K. Y. Camsari, and S. Datta, *Low-Barrier Nanomagnets as p-Bits for Spin Logic*, IEEE Magn. Lett. **8**, 4105305 (2017).

[23] K. Y. Camsari, R. Faria, B. M. Sutton, and S. Datta, *Stochastic P-Bits for Invertible Logic*, Phys. Rev. X **7**, 031014 (2017).

[24] R. Faria, J. Kaiser, K. Y. Camsari, and S. Datta, *Hardware Design for Autonomous Bayesian Networks*, Front. Comput. Neurosci. **15**, 584797 (2021).

[25] W. A. Borders, A. Z. Pervaiz, S. Fukami, K. Y. Camsari, H. Ohno, and S. Datta, *Integer Factorization Using Stochastic Magnetic Tunnel Junctions*, Nature **573**, 390 (2019).

[26] F. Heras, J. Larrosa, S. de Givry, and T. Schiex, *2006 and 2007 Max-SAT Evaluations: Contributed Instances*, J. Satisf. Boolean Model. Comput. **4**, 239 (2008).

[27] B. Molnár, M. Varga, Z. Toroczkai, and M. Ercsey-Ravasz, *A High-Performance Analog Max-SAT Solver and Its Application to Ramsey Numbers*, ArXiv:1801.06620 (2018).

[28] M. Kowalsky, T. Albash, I. Hen, and D. A. Lidar, *3-Regular 3-XORSAT Planted Solutions Benchmark of Classical and Quantum Heuristic Optimizers*, ArXiv:2103.08464v1 (2021).

[29] K. Boothby, P. Bunyk, J. Raymond, and A. Roy, *Next-Generation Topology of D-Wave Quantum Processors*, ArXiv:2003.00133 (2020).

[30] B. Molnár, F. Molnár, M. Varga, Z. Toroczkai, and M. Ercsey-Ravasz, *A Continuous-Time MaxSAT Solver with High Analog Performance*, Nat. Commun. **9**, 4864 (2018).

[31] F. L. Traversa, P. Cicotti, F. Sheldon, and M. Di Ventra, *Evidence of Exponential Speed-Up*

*in the Solution of Hard Optimization Problems*, Complexity **2018**, 7982851 (2018).

[32] K. Hayakawa, S. Kanai, T. Funatsu, J. Igarashi, B. Jinnai, W. A. Borders, H. Ohno, and S. Fukami, *Nanosecond Random Telegraph Noise in In-Plane Magnetic Tunnel Junctions*, Phys. Rev. Lett. **126**, 117202 (2021).

[33] C. Safranski, J. Kaiser, P. Trouilloud, P. Hashemi, G. Hu, and J. Z. Sun, *Demonstration of Nanosecond Operation in Stochastic Magnetic Tunnel Junctions*, Nano Lett. **21**, 2040 (2021).

[34] F. L. Traversa and M. Di Ventra, *Polynomial-Time Solution of Prime Factorization and NP-Complete Problems with Digital Memcomputing Machines*, Chaos An Interdiscip. J. Nonlinear Sci. **27**, 023107 (2017).

[35] J. Aiken and F. L. Traversa, *Memcomputing for Accelerated Optimization*, ArXiv:2003.10644v1 (2020).

[36] S. Prestwich, *CNF Encodings*, in *Handbook of Satisfiability* (IOS Press, 2008), pp. 73–95.

[37] N. A. Aadit, A. Grimaldi, M. Carpentieri, L. Theogarajan, J. M. Martinis, G. Finocchio, and K. Y. Camsari, *Massively Parallel Probabilistic Computing with Sparse Ising Machines*, ArXiv:2110.02481 (2021).

[38] P. J. M. van Laarhoven and E. H. L. Aarts, *Simulated Annealing*, in *Simulated Annealing: Theory and Applications* (Springer Netherlands, Dordrecht, 1987), pp. 7–15.

[39] R. H. Swendsen and J. S. Wang, *Replica Monte Carlo Simulation of Spin-Glasses*, Phys. Rev. Lett. **57**, 2607 (1986).

[40] D. J. Earl and M. W. Deem, *Parallel Tempering: Theory, Applications, and New Perspectives*, Phys. Chem. Chem. Phys. **7**, 3910 (2005).

[41] C. Blum and A. Roli, *Metaheuristics in Combinatorial Optimization*, ACM Comput. Surv. **35**, 268 (2003).

[42] L. Lopez-Diaz, L. Torres, and E. Moro, *Transition from Ferromagnetism to Superparamagnetism on the Nanosecond Time Scale*, Phys. Rev. B **65**, 224406 (2002).

[43] J. Cai, B. Fang, L. Zhang, W. Lv, B. Zhang, T. Zhou, G. Finocchio, and Z. Zeng, *Voltage-Controlled Spintronic Stochastic Neuron Based on a Magnetic Tunnel Junction*, Phys. Rev. Appl. **11**, 034015 (2019).

[44] G. Finocchio, T. Moriyama, R. De Rose, G. Siracusano, M. Lanuzza, V. Puliafito, S. Chiappini, F. Crupi, Z. Zeng, T. Ono, and M. Carpentieri, *Spin–Orbit Torque Based Physical Unclonable Function*, J. Appl. Phys. **128**, 033904 (2020).

[45] G. Bertotti, I. D. Mayergoyz, and C. Serpico, *Analytical Solutions of Landau-Lifshitz Equation for Precessional Dynamics*, Phys. B Condens. Matter **343**, 325 (2004).

[46] G. Bertotti, I. D. Mayergoyz, and C. Serpico, *Analysis of Random Landau-Lifshitz Dynamics by Using Stochastic Processes on Graphs*, J. Appl. Phys. **99**, 10 (2006).

[47] A. Romeo, G. Finocchio, M. Carpentieri, L. Torres, G. Consolo, and B. Azzerboni, *A Numerical Solution of the Magnetization Reversal Modeling in a Permalloy Thin Film Using Fifth Order Runge–Kutta Method with Adaptive Step Size Control*, Phys. B Condens. Matter **403**, 464 (2008).

[48] W. F. Brown, *Thermal Fluctuations of a Single-Domain Particle*, Phys. Rev. **130**, 1677 (1963).

[49] G. Finocchio, I. N. Krivorotov, X. Cheng, L. Torres, and B. Azzerboni, *Micromagnetic Understanding of Stochastic Resonance Driven by Spin-Transfer-Torque*, Phys. Rev. B **83**, 134402 (2011).

[50] G. Bertotti, C. Serpico, I. D. Mayergoyz, A. Magni, M. D'Aquino, and R. Bonin, *Magnetization Switching and Microwave Oscillations in Nanomagnets Driven by Spin-Polarized Currents*, Phys. Rev. Lett. **94**, 127206 (2005).